# Effects of Electron-Beam Irradiation on Graphene Oxide

P. Adamson and S. Williams

*Department of Physics and Geosciences, Angelo State University, ASU Station #10904, San Angelo, Texas 76909*

**Abstract.** Graphene oxide (GO) is a nanofilm composed of graphene with various oxygen functional groups attached. GO is of interest due to its unique mechanical-enhancement properties, its tunable electronic properties, and its potential use in the wide-scale production of graphene. Scanning electron microscopes (SEMs) are frequently used to characterize and study GO films. The purpose of this project was to study the effects of SEM-imaging on GO films. Using an SEM, we irradiated GO samples at electron beam-energies of 10, 20, and 30 keV (at a constant emission current of ~40 ± 2 μA) for times ranging from 15 minutes to one hour. Raman D- and G-band intensities were used to examine structural modifications/damage to GO samples as a function of beam energy and exposure time. The results suggest that imaging with a 30 keV electron beam for 30 minutes may lead to the formation of amorphous carbon, while imaging with 10 keV or 20 keV beams for 30 minutes does not have a significant effect on GO samples.

## INTRODUCTION

Graphene is a single atomic layer of carbon atoms arranged in a hexagonal form. Graphene oxide (GO) is a sheet of graphene with various oxygen functional groups covalently attached. GO has attracted substantial interest due to its unique mechanical properties[1] (which include the ability to enhance the tensile strengths of materials[2,3]), its tunable electronic properties[4], and its potential use in the wide-scale production of graphene[5]. Scanning electron microscopes (SEMs) are commonly used to analyze the surface morphology[1] of GO and graphene samples.

SEMs produce images by scanning samples' surfaces with electron beams with energies typically ranging from 500 eV to 50 keV. Exposure of GO and other nanomaterials to electron beams during SEM-characterization is known to introduce defects[7] that can significantly alter thermal and electrical conduction[6].

Raman spectroscopy is a vibration-based technique that is commonly used to characterize disorder in $sp^2$ carbon materials[8]. Raman spectroscopy involves shining monochromatic light (typically from a laser) on a sample and studying the light that is inelastically scattered from it in order to learn about the vibrational and rotational modes of the sample. The Raman D-band is commonly found in Raman spectra at a wavenumber of ~1350 cm$^{-1}$. The prominent D-band peak (which is not present in the Raman spectra of pristine graphene) is defect-dependent. In the case of GO, the D-band's existence in Raman spectra is primarily due to structural imperfections as the result of the presence of hydroxyl and epoxide groups. The first-order G-band is commonly found at a wavenumber of ~1580 cm$^{-1}$. It is related to the stretching of $sp^2$ pairs, and, unlike the D-band, is not defect-dependent. The Raman D- to G-band peak-intensity ratio ($I_D/I_G$) is, thus, proportional to the $sp^2$ defect-density of carbon-based samples[9]. In this study, we used Raman analysis to compare defect-densities in GO samples irradiated using an SEM and as-purchased GO samples.

## EXPERIMENTAL PROCEDURE

For this experiment we used GO samples (Graphenea Inc., Spain) prepared by the filtration of a monolayer GO dispersion. The circular GO samples had thicknesses of 12-15 μm and diameters of 4.5 ± 0.2 mm. The GO films were composed of 49-56 wt% carbon, 41-50 wt% oxygen, 0-2 wt% sulfur, 0-1 wt% nitrogen, and 0-1 wt% hydrogen.

Samples were irradiated using a Hitachi S-3000N SEM at emission currents of 40 ± 2 μA and accelerating voltages ranging from 10 kV to 30 kV. While irradiating the GO samples, every effort was made to stabilize the SEM's emission current at 40 μA by continuously monitoring the emission current and adjusting the filament voltage, when necessary. As the emission current was essentially held constant during all irradiations, and as the same aperture and magnification settings were used during each irradiation, irradiation dosages were approximately proportional to exposure times. An SEM image of a typical GO sample is shown in Fig. 1.

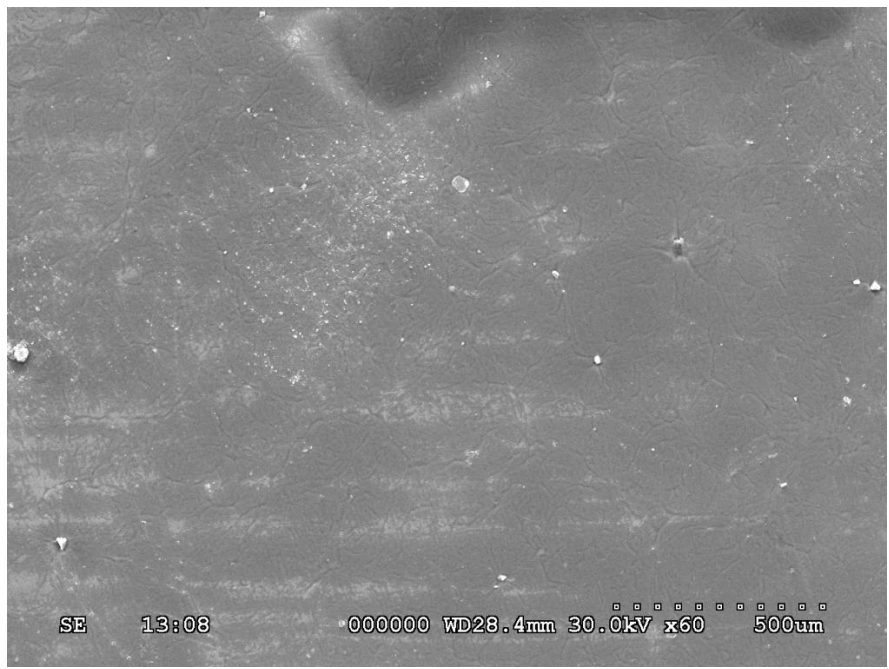

**FIGURE 1.** SEM image of typical GO sample.

Raman spectra were obtained using a Bruker Optics Senterra dispersive Raman microscope spectrometer, equipped with a 2 mW, 532 nm-wavelength laser. Background contributions were subtracted from the spectra using software developed by Candeloro *et al.*[10] A typical Raman spectrum from an irradiated GO sample (with background subtracted) is shown in Fig. 2.

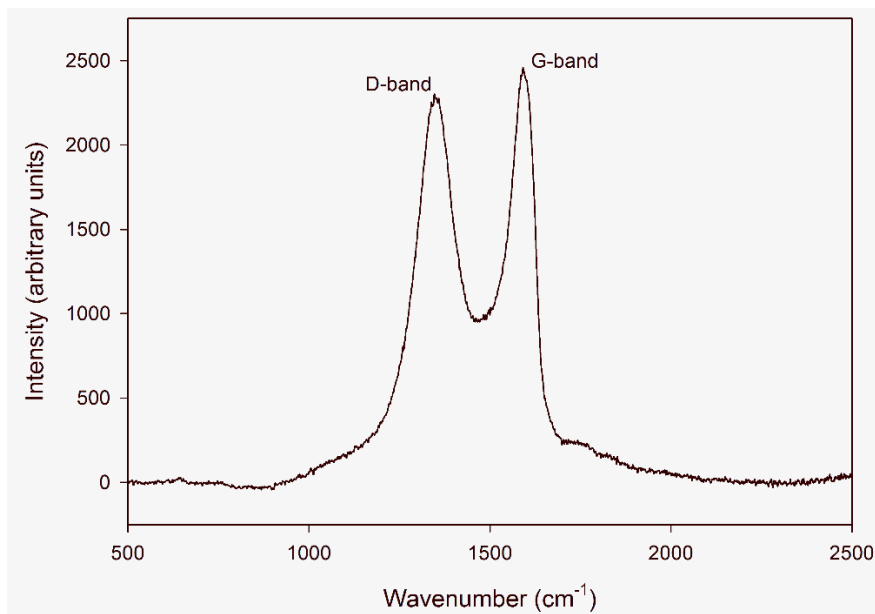

**FIGURE 2.** Typical Raman spectrum from an irradiated GO sample after background has been subtracted.

## RESULTS AND DISCUSSION

The $I_D/I_G$ values for GO films irradiated using 0 keV, 10 keV, 20 keV, and 30 keV electron beams for 30 minutes are shown in Fig. 3. (The sample "irradiated" using a 0 keV beam is simply an as-purchased sample.) The uncertainties represented by the error bars in Fig. 3 were calculated by combining the uncertainties in background subtraction and statistical uncertainties in quadrature. The data shown in Fig. 3 suggests that $I_D/I_G$ decreases as the SEM's electron beam-energy increases. This is likely due to the introduction of defects leading to the formation of amorphous carbon. As the beam increases the number of defects in the GO and reduces the in-plane correlation length, the number of ordered rings in the GO decreases, which leads to a decrease in $I_D$[7]. As the G-band is only related to the stretching of $sp^2$ pairs, its intensity remains unchanged. However, the wavenumber at which the peak-intensity of the G-band ($I_G$) appears was observed to consistently shift to lower wavenumbers as the electron beam-energy was increased. The wavenumbers at which $I_G$ appears in the Raman spectra for GO samples irradiated with 0, 10, 20, and 30 keV electrons are shown in Table 1. This shift is also consistent with a partial change from a crystalline to amorphous structure[7]. The results shown in Fig. 3 are consistent with the results of similar experiments involving graphene performed by Teweldebrhan and Balandin[6].

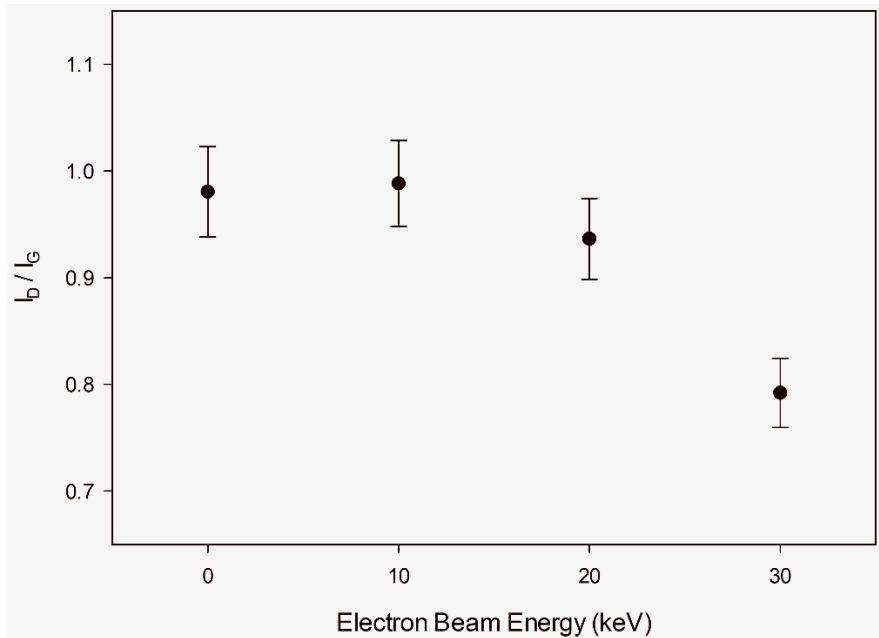

**FIGURE 3.** Raman D- to G-band peak-intensity ratios ($I_D/I_G$) for GO samples irradiated for 30 minutes using 10 keV, 20 keV, and 30 keV electron beams. The data point at 0 keV corresponds to an as-purchased (unirradiated) GO sample.

| Accelerating Voltage (kV) | $I_G$ Wavenumber (cm$^{-1}$) |
|---|---|
| 0 | 1594.5 |
| 10 | 1592.0 |
| 20 | 1591.5 |
| 30 | 1591.0 |

**TABLE 1.** Wavenumbers at which IG appears in the Raman spectra for GO irradiated with 0, 10, 20, and 30 keV electrons.

The $I_D/I_G$ values for GO films irradiated using a 30 keV electron beam for 0, 15, 30, 45, and 60 minutes are shown in Fig. 4. Just as in Fig. 3, the uncertainties represented in Fig. 4 were calculated by combining uncertainties in background subtraction and statistical uncertainties in quadrature. The decrease in $I_D/I_G$ values between 0 and 30 minutes of irradiation-time may be due to the aforementioned change from a crystalline to amorphous structure. However, the increase in $I_D/I_G$ values after 30 minutes of irradiation-time is not completely understood. The results shown in Fig. 4. are not consistent with those of Teweldebrhan and Balandin[6]. However, the experiments performed by Teweldebrhan and Balandin involved graphene, rather than GO, and it is possible that discrepancies in the results are due to the presence of oxygen groups. It is also possible that the uncertainty in some factor, such as background subtraction, was underestimated, and that the low $I_D/I_G$ value for 30 minutes of irradiation-time is simply the result of experimental error.

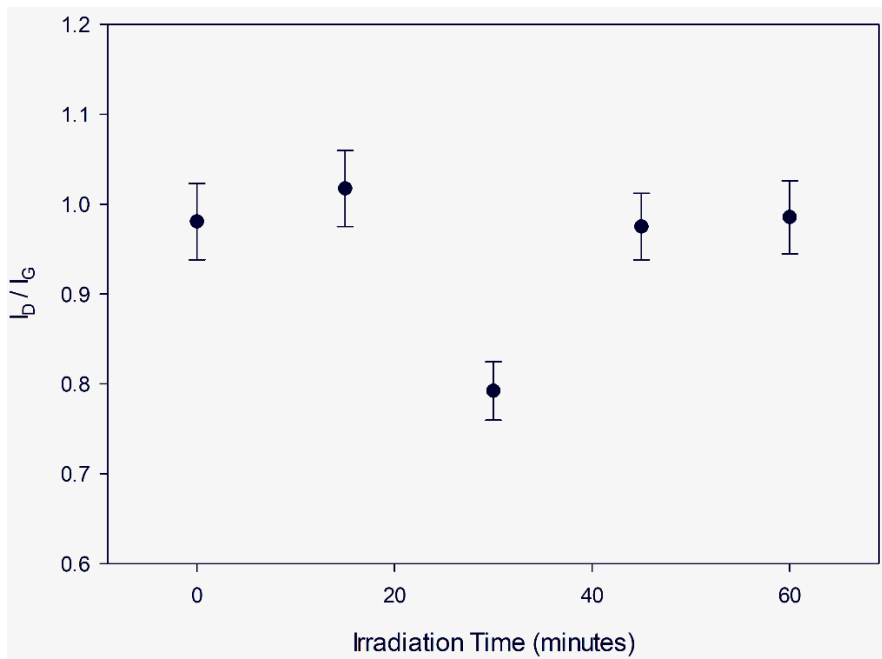

**FIGURE 4.** Raman D- to G-band peak-intensity ratios ($I_D/I_G$) for GO samples irradiated using a 30 keV beam for 0, 15, 30, 45, and 60 minutes.

## CONCLUSIONS

There is still a need for measurements of $I_D/I_G$ values at more frequent intervals to better understand the effects of irradiation-time on GO samples. Presently, we do not have enough information to completely understand the effects of long-term electron irradiation-times on GO films.

The data presented here suggests that 30 minutes of irradiation-time has little effect on GO for electron beam-energies of 10 keV and 20 keV. However, exposure to a 30 keV electron beam for a duration of 30 minutes led to a partial change from a crystalline to amorphous structure. This change is evident through a decrease in $I_D/I_G$ values, as well as through a consistent shift in the peak-intensity of the G-band to lower wavenumbers, as the electron beam-energy was increased. In the future, additional experiments should be performed in order to more accurately determine the accelerating voltage at which amorphous carbon begins to form as the result of electron-beam irradiation. For the time being, the results presented here should be taken into consideration by researchers when performing analysis of GO samples subsequent to SEM imaging.

## ACKNOWLEDGMENTS

The authors wish to thank Juliusz Warzywoda at Texas Tech University for his assistance with obtaining the Raman spectra.